# Induced spin-orbit coupling in silicon thin films by bismuth doping

F.Rortais[1,*], S.Lee[1], R.Ohshima[1], S.Dushenko[1], Y.Ando[1] and M.Shiraishi[1]

[1]Department of Electronic Science and Engineering, Kyoto University, Kyoto, Kyoto 615-8510

E-mail: rortais.fabien.85z@st.kyoto-u.ac.jp

We demonstrate an enhancement of the spin-orbit coupling in silicon (Si) thin films by doping with bismuth (Bi), a heavy metal, using ion implantation. Quantum corrections to conductance at low temperature in phosphorous-doped Si before and after Bi implantation is measured to probe the increase of the spin-orbit coupling, and a clear modification of magnetoconductance signals is observed: Bi doping changes magnetoconductance from weak localization to the crossover between weak localization and weak antilocalization. The elastic diffusion length, phase coherence length and spin-orbit coupling length in Si with and without Bi implantation are estimated, and the spin-orbit coupling length after the Bi doping becomes the same order of magnitude ($L_{so}$ = 54 nm) with the phase coherence length ($L_{\varphi}$ = 35 nm) at 2 K. This is an experimental proof that the spin-orbit coupling strength in Si thin film is tunable by doping with heavy metals.

Silicon (Si) is an attracting material for spintronics[1–11] because its spin-orbit coupling is intrinsically weak, and spin lifetime in it can be longer than one nanosecond[2,10] even at room temperature. Almost all the recent spintronics studies of Si used lateral spin valves for spin injection, detection and transport[2–10]. The difficulty and low efficiency of spin injection into semiconductor has been one of the central issues in semiconductor spintronics: the so-called conductivity mismatch[12] hinders the spin injection from ferromagnetic metals into the semiconductor channel. Although insertion of tunneling barrier between ferromagnetic metals and semiconductors allows to inject spins into semiconductors, spin injection/detection using the ordinary and the inverse spin Hall effects (SHE and ISHE)—instead of electrical spin injection from ferromagnetic metals—can be the other pathway to avoid the conductivity mismatch[13–18].

Si possesses a low spin-orbit coupling because of the small electric charge of its nucleus and a lack of an internal electric field due to the lattice inversion symmetry, which allows devices with spin transport but limits new device-designing possibilities with the spin-charge conversion effects—the SHE and the ISHE mentioned above. However, if it can be engineered to possess a significant spin-orbit coupling, the SHE and the ISHE in Si can be used for spin injection/detection. Such additional functionality will pave the way to create all-Si spin devices consisting of injector, detector and transport medium made of Si—a breakthrough for semiconductor spintronics and significant advance in applied physics. The purpose of this study is to create a sizable spin-orbit interaction in Si by implantation of a heavy element, bismuth (Bi). To probe the strength of the spin-orbit coupling, quantum corrections to the conductivity at low temperatures were measured in Si with and without Bi implantation. The enhancement of the spin-orbit interaction is discussed using the spin-orbit coupling length extracted from the measurements.

Samples were fabricated using an intrinsic silicon-on-insulator (SOI) substrate with a 100 nm-thick Si channel isolated from the backside bulk Si by 200 nm-thick layer of $SiO_2$. The Si channel was doped with phosphorus (P) by ion implantation to assure ohmic contacts for electrical measurements. The doses and energies of ion beam necessary to reach a homogeneous profile of $1\times10^{20}\,cm^{-3}$—which exceeds the degenerate limit of $3.5\times10^{18}\,cm^{-3}$ for P-doped Si[19]—were calculated from the ion implantation profile simulated with the Transport of Ions in Matter (TRIM) program. The three implantation steps are summarized in Table I, and the simulated implantation profile is shown in Fig. 1(a). To activate P and decrease the number of defects in the Si channel induced by the implantation, the substrate was annealed using rapid thermal annealing at 500° C for 10 s for the first step, and at 900° C for 1 s for the last step, with a 4 s temperature ramp up in-between. A part of the substrate was used as reference samples (SOI:P), while the rest was additionally doped by Bi to form a double donor system (SOI:P:Bi). Energies and doses for Bi implantation were chosen to realize homogeneous doping concentration of $5.0 \times 10^{19}$ cm$^{-3}$ and are summarized in Table I. The doping profile simulated by TRIM is shown in Fig. 1(a) (blue-filled squares). It was previously demonstrated that such doping concentration is above the critical level for metal-semiconductor transition for Si doped with Bi[20]. The implantation was followed by a thermal annealing at 600° C for 30 min under argon atmosphere to electrically activate the Bi dopants[21].

Table I. Energies and doses used for P and Bi doping.

| Step n° | P doping | | Bi doping | |
|---|---|---|---|---|
| | Energy (keV) | Dose ($\times 10^{13}$ cm$^{-2}$) | Energy (keV) | Dose ($\times 10^{13}$ cm$^{-2}$) |
| 1 | 5 | 8 | 35 | 5.4 |
| 2 | 15 | 13 | 70 | 7.4 |
| 3 | 35 | 50 | 120 | 11.0 |
| 4 | | | 250 | 19.0 |

The electron beam lithography and argon milling processes were used to pattern the Si channel into Hall bar devices as shown in FIG. 1(b). It should be noted that the Bi concentration is very low in the top 20 nm of Si channel (see Fig. 1(a)) and, therefore, we etched the top 50 nm to use the uniformly doped region as a channel. The etching was performed on both types of samples (SOI:P and SOI:P:Bi) to ensure the same channel structure. The Hall bars were 20 μm in width and 110 μm in length in-between the voltage measurement branches (terminals 3 and 5 in Fig. 1(b)). The electrical contacts on top of the Si channel were formed in the two-step process: deposition of 30 nm-thick AuSb followed by annealing for 30 s at 300° C, and then deposition of Ti(3 nm)/Au(70 nm) bilayer.

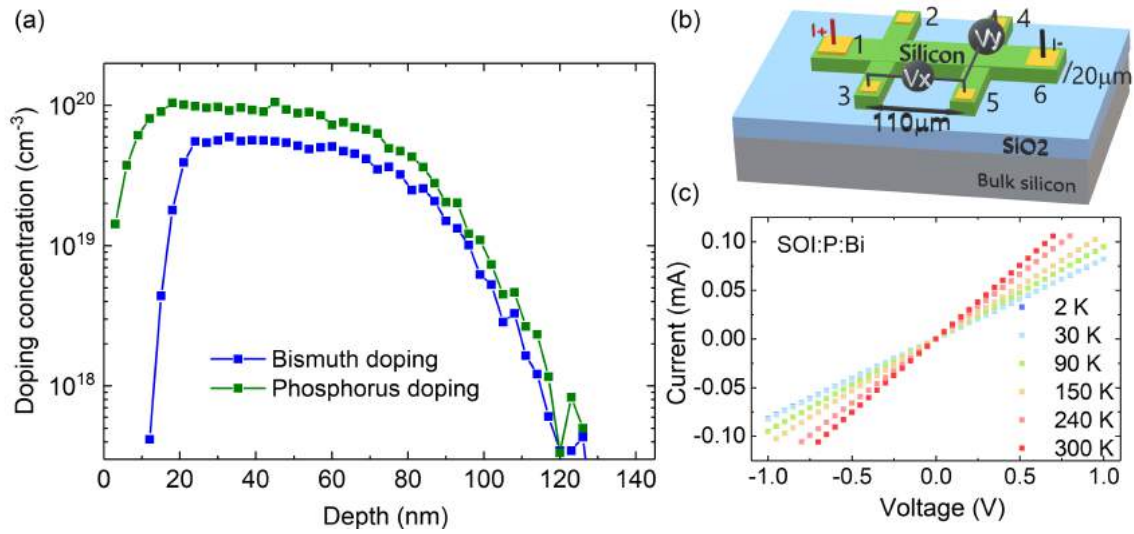

FIG. 1. (a) Doping profiles of Bi and P in Si simulated using the Transport of Ions in Matter (TRIM) program with the irradiation beam parameters specified in Table I. Solid lines are guide to the eye. (b) Schematic layout of the Hall bar devices used to carry out measurements. (c) Temperature dependence of $I$-$V$ characteristics measured between injection contacts (terminals 1 and 6) of SOI:P:Bi sample. Sample exhibited linear $I$-$V$ characteristics in the whole measured temperature range.

Figure 1(c) demonstrates the linear $I$-$V$ characteristics measured from the SOI:P:Bi sample, confirming successful fabrication of the ohmic contacts with the Si channel. The linear $I$-$V$ characteristics were measured in the whole temperature range from 300 K down to 2 K, which rules out contribution of non-linearities in $I$-$V$ characteristic on the magnetoconductance measurements at low temperatures. Figure 2(a) shows the temperature dependence of resistivity for the SOI:P and the SOI:P:Bi samples. For the SOI:P, the resistivity increased from $1.3 \times 10^{-3}$ Ω·cm at 2 K to $2.4 \times 10^{-3}$ Ω·cm at 300 K. The metallic behavior confirms the degenerate state of the Si channel after the P doping. On the contrary, in the Bi-doped sample (SOI:P:Bi), a decrease of resistivity from $8.2 \times 10^{-3}$ Ω·cm at 2 K to $4.2 \times 10^{-3}$ Ω·cm at 300 K was observed. To clarify its origin, we measured the Hall effect to extract carrier concentration in the channel. Consequently, the carrier concentrations at 2K were estimated to be $4.0 \times 10^{19}$ cm$^{-3}$ and $5.0 \times 10^{19}$ cm$^{-3}$ for the SOI:P and the SOI:P:Bi, respectively. As expected, the carrier concentration increased after the Bi doping in SOI:P:Bi samples compared with the SOI:P samples. However, the measured carrier densities for both P and Bi are smaller than those predicted by the TRIM simulations, which indicates that only a part of the implanted ions is electrically activated. Figure 2(b) shows the mobilities of the Si channel extracted from the Hall effect measurements. The carrier mobilities were strongly reduced after Bi doping from 118 cm$^2$·V$^{-1}$·s$^{-1}$ (SOI:P) to 14 cm$^2$·V$^{-1}$·s$^{-1}$ (SOI:P:Bi) at 2 K and from 64 cm$^2$·V$^{-1}$·s$^{-1}$ (SOI:P) to 18 cm$^2$·V$^{-1}$·s$^{-1}$ (SOI:P:Bi) at 300 K. This decrease of the mobilities is attributed to the increased number of scattering centers in the Si channel with the implantation of Bi ions and the additional crystal defects produced during the implantation, as was also reported in the previous studies[22–24]. However, we stress that despite the decrease in the mobilities, linear $I$-$V$ characteristics were still obtained for SOI:P and SOI:P:Bi samples at all temperatures.

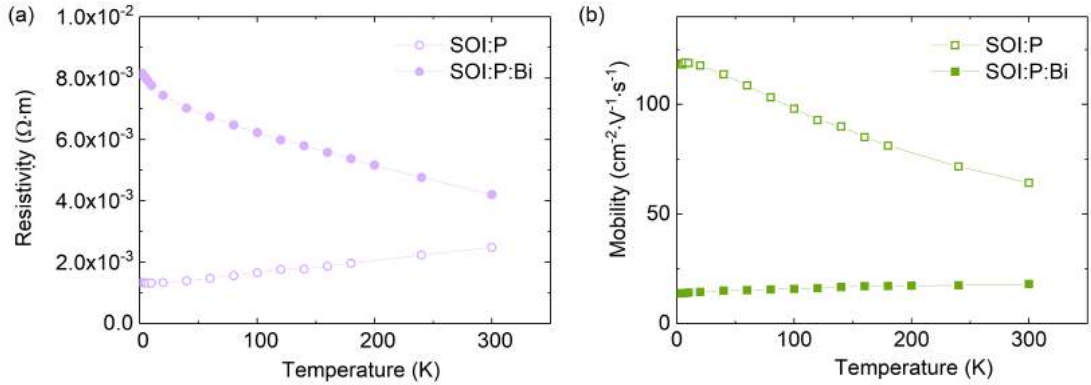

FIG. 2. Temperature dependence of (a) resistivity and (b) mobility of the P-doped Si channel without Bi doping (SOI:P, open symbols) and with Bi doping (SOI:P:Bi, filled symbols). Solid lines are guide to the eye.

To compare the spin-orbit coupling strength in the Si channel with and without Bi doping (samples SOI:P:B and SOI:P, respectively), we measured quantum corrections to the conductance. At low temperature, a conduction electron can be scattered by impurities several times without losing its phase coherence, which creates interference on self-crossing paths and leads to localization of the wave function (i.e. decreased conductance)—the phenomenon known as weak localization (WL)[25–28]. The phase coherence length, $L_\varphi$, defines the characteristic length during which the phase is conserved. The spin-orbit coupling—which can be characterized by spin-orbit coupling length $L_{so}$—also influences the relative phase between interfering waves by coupling spin of the electron to its momentum. This results in destructive interference on the self-crossing paths—the weak antilocalization effect (WAL) that manifests itself in the increased conductance [27,26]. Application of a magnetic field introduces an additional phase factor that destroys interference and leads to suppression of the conductance change due to the WL and WAL effects. Thus, both effects result in magnetoconductance, however, its polarity is reversed between systems with a strong spin-orbit coupling (WAL case) and a week spin-orbit coupling (WL case)[26,29–31]. The crossover between the WAL and the WL is governed by the ratio $L_\varphi/L_{so}$[27]—making magnetoconductance a sensitive tool to determine the spin orbit coupling strength[27], as was already demonstrated in various systems [30–34].

The magnetoconductance measurements were carried out using Physical Property Measurement System (PPMS, Quantum Design): the longitudinal voltage between terminals 3 and 5 (see FIG. 1(b)) was measured under an application of the out-of-plane magnetic field and the injection current of 10 μA. The linear background, which probably originates from the thermal drift, was subtracted from the data before further processing. We calculated the normalized conductance $\Delta G/G_0` = (G(B)-G(B=0))/G_0`$, where $G(B)$ is the conductance under the application of the magnetic field $B$; $G(B=0)$ is the zero-magnetic field conductance; and $G_0` = G_0/(2\pi) = e^2/(2\pi^2 \hbar)$, where $G_0$ is conductance quantum, $e$ is the elementary charge, $\hbar$ is the reduced Planck constant. Figures 3(a) and (b) show the normalized conductance as a function of the magnetic field for the SOI:P and the SOI:P:Bi samples, respectively. The sample without Bi doping (SOI:P) shows increasing of the conductance with increasing the magnetic field—a signature of the WL effect. The amplitude of the quantum corrections to the conductance by the WL can be extracted by removing the parabolic contribution due to the classical magnetoconductance (not shown). The WL signal monotonically increased from 20 K to 2 K, with an amplitude of 1.18±0.04% of the zero-field conductance at 2 K. Our results confirm previous observation of the WL in P-doped Si[10]. The sample with Bi doping (SOI:P:Bi) exhibited strikingly different behavior of magnetoconductance. The WL in SOI:P:Bi sample at 2 K was more than five times smaller than in SOI:P sample without Bi doping. Furthermore, the amplitude of the WL decreased from 5 K (0.28±0.01%) to 2 K (0.22±0.01%) in the SOI:P:Bi, in contrast to the increase from 5 K (0.88±0.04%) to 2 K (1.18±0.04%) in the SOI:P. The change of the shape of the magnetoconductance curve around the zero magnetic field from narrow dip (Fig. 3(a)) to broad dip (Fig. 3(b))—along with the strong decrease in the amplitude of the WL—indicates transition from the WL to the WAL in the SOI:P:Bi sample. Such behavior is expected from magnetoconductance of a material that possesses a significant spin-orbit coupling, leading to the spin-orbit coupling length and the phase coherence lengths to have the same order of magnitude[27]. The characteristics lengths $L_{so}$ and $L_\varphi$ can be extracted from the magnetoconductance data by using the Hikami-Larkin-Nagaoka model[35]:

$$\frac{\Delta G}{G_0} = \frac{G(B) - G(0)}{\frac{e^2}{2\pi^2\hbar}}$$
$$= \left[\ln\left(\frac{B_\phi}{B}\right) - \psi\left(\frac{1}{2} + \frac{B_\phi}{B}\right)\right] + 2\left[\ln\left(\frac{B_{SO} + B_e}{B}\right) - \psi\left(\frac{1}{2} + \frac{B_{SO} + B_e}{B}\right)\right]$$
$$- 3\left[\ln\left(\frac{\frac{4}{3}B_{SO} + B_\phi}{B}\right) - \psi\left(\frac{1}{2} + \frac{\frac{4}{3}B_{SO} + B_\phi}{B}\right)\right] + pB^2, \quad (1)$$

where $\Psi$ is the digamma function defined as: $\Psi(x) = \Gamma'(x)/\Gamma(x)$ with $\Gamma(x) = \int_0^\infty y^{x-1}e^{-y}\,dy$; the characteristic fields $B_\varphi$, $B_{so}$ and $B_e$ are connected to the characteristic lengths $L_\varphi$, $L_{so}$ and $L_e$, respectively [27] by $B_i = \frac{\hbar}{4eL_i^2}$, where the suffix $i$ stands for $\varphi$, SO, $e$; $p$ is the coefficient describing strength of the parabolic contribution due to the classical magnetoconductance. We stress that the classical magnetoconductance was considered even at low magnetic field: it was evaluated using magnetoconductance data at the magnetic field $|B| > 2$ T at each temperature and then included as a fixed parameter in the fitting function (Eq. (1)). The resulting fitting curves are plotted as black solid lines in Figs. 3(a) and (b). The elastic diffusion length, phase coherence length and spin-orbit coupling length extracted from the fitting are summarized in Figs. 3(c) and (d) for the SOI:P and the SOI:P:Bi. The elastic diffusion length has the same order of magnitude (10 nm-25 nm) in the samples with and without Bi doping. The phase coherence length, however, decreased after the Bi doping (with the strongest decrease from 177 nm to 35 nm at 2 K), indicating that Bi atoms in the Si are mainly acting as inelastic scattering centers. The use of the Hikami-Larkin-Nagaoka model—a 2D model—is justified because the extracted phase coherence length has the same order of magnitude or is superior to the thickness of the Si channel. The fitting results for the SOI:P:Bi demonstrated that Bi doping is capable of inducing the spin-orbit coupling length of the same order of magnitude with the phase coherence length ($L_{so}$= 54 nm and $L_\varphi$= 35 nm at 2K).

As expected, the implantation of heavy elements like Bi in the Si channel as new scattering centers decreased the phase coherence length. This implantation-induced disorder in the Si channel can be seen from the modification of the mobility and its temperature dependence (Fig. 2(b)). The temperature dependence of the phase coherence length itself was also modified by the Bi doping (FIG. 3(c) and (d)). In the temperature range from 2 K to 5 K, the reference SOI:P sample presents a decrease of 46%, while the Bi doped sample, the SOI:P:Bi, shows only 29% decrease and a less steeper decreasing law, which is characteristic to a different scattering mechanism[36]. The ratio $B_{so}/B_\varphi \propto L_\varphi^2/L_{so}^2$ that describes the crossover between the WL/WAL[27] is modified by the addition of Bi: the increased spin-orbit coupling strength led to the observed crossover in the magnetoconductance close to the zero magnetic field. The crossover between the WL and the WAL was observed—rather than full polarity reversal of magnetoconductance correction—since the spin-orbit coupling length ($L_{so}$= 54 nm at 2K) is still higher than phase coherence length (35 nm), as was also shown in the previous studies[27]. Increasing number of electrically activated Bi atoms can provide a way to further enhance the spin-orbit coupling without increase in Bi doping concentration.

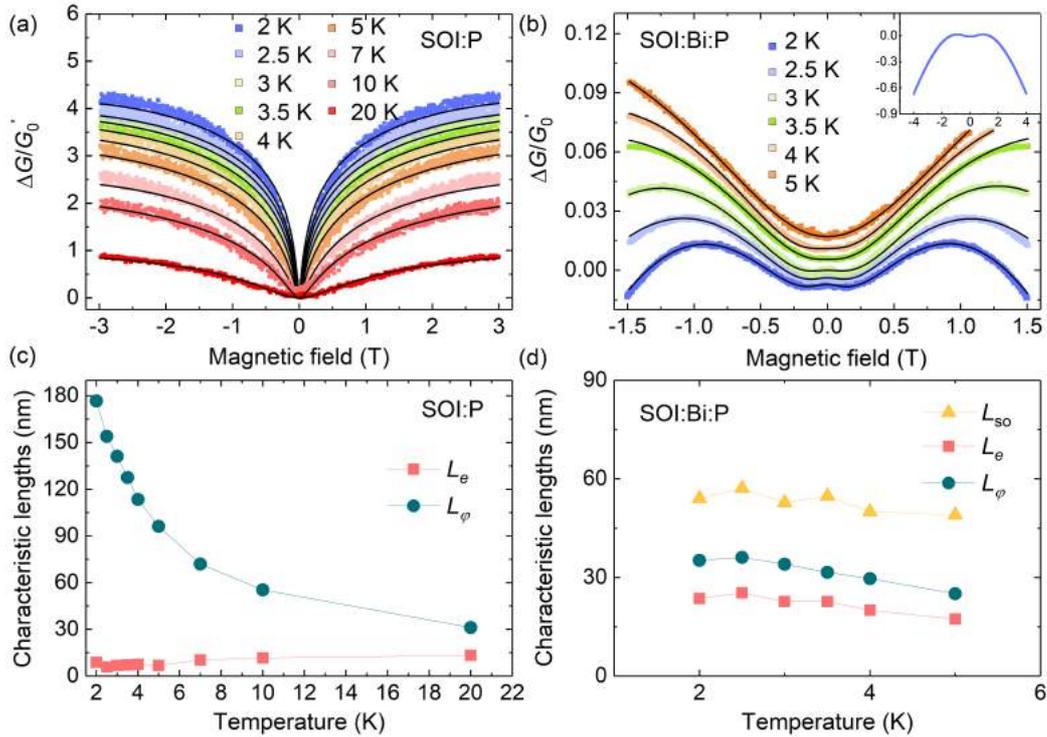

FIG. 3. (a) The normalized magnetoconductance $\Delta G/G_0` = (G(B)-G(B=0))/G_0`$ of (a) P-doped Si channel without Bi doping (SOI:P) and (b) P-doped Si channel with Bi doping (SOI:P:Bi) measured at different temperatures. Inset in panel (b) shows full magnetic field $B$ scan range data measured at $T$=2 K. The curves are offset along the $y$-axis for clarity. Solid black lines show fitting by the Eq. (1), as described in the text. Temperature dependence of the mean free path length $L_e$ (squares), the phase coherence length $L_\varphi$ (circles) and the spin-orbit coupling length $L_{so}$ (triangles) of (c) P-doped Si channel without Bi doping (SOI:P) and (d) P-doped Si channel with Bi doping (SOI:P:Bi). Solid lines are guide to the eye.

For spintronics applications, one of the key parameters that characterizes material is the spin diffusion length—the mean distance between spin-flipping events. Long spin diffusion length characterizes how far spin can carry information and is crucial for device application. The exact relation between the spin diffusion length and the spin-orbit coupling length in Si is complicated because of the anisotropic band structure[34,37,38]. In addition, in Si with Bi doping, the effective mass theory is expected to be less applicable because of the large ionization energy and small Bohr radius[22]. Furthermore, the temperature dependence of the phase coherence length is modified by the Bi implantation—a proof of the modification of the scattering mechanism—which complicates the comparison between the spin life time in two systems. However, the appearance of the WAL and the observable spin-orbit coupling length in Bi-doped samples signifies the enhancement of the spin-orbit coupling in Si, which inevitably leads to the decrease in the spin diffusion length. In this sense, the increased spin-orbit coupling provides an advantage of utilizing a spin-charge conversion mechanism, but, at the same time, introduces a new limitation on the size of spintronics devices.[39] The interplay between these two effects can be controlled by tuning Bi doping concentration—a pathway to create a device that integrates spin-charge conversion and spin transport.

In conclusion, we demonstrated a control over the strength of the spin-orbit coupling in Si channel using Bi doping. Magnetoconductance measurements were carried out on the Hall-bar-shaped P-doped Si devices with and without Bi doping. Strong modification of electrical transport properties after Bi doping—an increase in resistivity despite increased carrier concentration and decrease of mobility and modification of its temperature behavior—indicated enhanced impurity scattering. Quantum corrections to conductance were strongly modified by Bi doping: a WL signal of 1.18±0.04% at 2 K was suppressed to 0.22±0.01% after the doping. Also the temperature dependence of WL amplitude is reversed between 2 K and 5 K for SOI:P and SOI:P:Bi respectively. Moreover, close to the zero magnetic field, we observed a beginning of polarity reversal of magnetoconductance in Bi-doped samples—a signature of a significant spin-orbit coupling in our Si devices. By using the magnetoconductance fitting by the Hikami-Larkin-Nagaoka model, we demonstrated a spin-orbit coupling length of 54 nm at 2 K in SOI:P:Bi. Our result paves the way for control over the spin-orbit coupling in Si by using doping with heavy elements, which opens new possibilities in designing Si spintronics devices with enhanced spin-orbit coupling enabling a use of spin-charge conversion via the spin Hall effects.

F.R. acknowledges support by JSPS (Japan Society for the Promotion of Science) Postdoctoral Fellowship and JSPS KAKENHI Grant No. 820170900047. A part of this study is supported by Grant-in-Aid for Scientific Research from the Ministry of Education, Culture, Sports, Science and Technology (MEXT) of Japan (Innovative Area "Nano Spin Conversion Science" No. 26103002 and Grant-in-Aid for Scientific Research (S) "Semiconductor Spincurrentronics" No. 16H06330).